\documentclass[aps,prd,reprint,superscriptaddress,amsmath,amssymb,nofootinbib,floatfix]{revtex4-2}
\usepackage{graphicx}
\usepackage{dcolumn}
\usepackage[utf8]{inputenc}

\usepackage{ulem}
\usepackage{xcolor}
\usepackage[unicode=true,colorlinks,linkcolor=blue,anchorcolor=blue,urlcolor=blue,citecolor=blue,breaklinks=true]{hyperref}
\usepackage{orcidlink}
\usepackage{subfigure}

\newcommand{\ii}{\mathrm{i}\,}

\newcommand{\pararrow}{\mathord{\buildrel{\lower3pt\hbox{$\scriptscriptstyle\leftrightarrow$}}\over {\partial}}} 
\newcommand{\pararrowk}[1]{\mathord{\buildrel{\lower3pt\hbox{$\scriptscriptstyle\leftrightarrow$}}\over {\partial}\hspace*{-0.18em}{}^#1}\hspace*{-0.18em} \,} 

\usepackage{physics}
\usepackage{bm}

\newcommand{\qfnu}{\affiliation{College of Physics and Engineering, Qufu Normal University, Qufu 273165, China}}

\newcommand{\itp}{\affiliation{CAS Key Laboratory of Theoretical Physics, Institute of Theoretical Physics, Chinese Academy of Sciences, Beijing 100190, China}}

\begin{document}

\title{ Hidden charmed decays of $X(3872)$ within the $D{\bar D}^*$ molecular framework }
    
    \author{Hao-Dong Cai\,\orcidlink{0009-0000-5269-8621}} \qfnu
    \author{Zhao-Sai Jia\,\orcidlink{0000-0002-7133-189X}} \qfnu \itp
	\author{Gang Li\,\orcidlink{0000-0002-5227-8296}} \email{gli@qfnu.edu.cn}\qfnu
	\author{Shi-Dong Liu\,\orcidlink{0000-0001-9404-5418}} \email{liusd@qfnu.edu.cn}\qfnu

\begin{abstract}
The dipionic transition of the $X(3872)$ to the $\eta_c$ was investigated using an effective Lagrangian approach. In this study, the $X(3872)$ was assumed to be a $D\bar{D}^\ast + \text{c.c.}$ bound state with the quantum numbers $J^{PC}=1^{++}$ and to decay via triangle and box loops. It is found that the partial decay widths arising from the box loops are one order of magnitude greater than those from the triangle ones. The decay widths are model-dependent, as characterized by the cutoff parameter introduced in the form factor. Moreover, the charged and neutral $D\bar{D}^\ast$ configurations of the $X(3872)$ and its mass also influence the decay widths. With our model parameters, the decay widths for the $X(3872)\to\pi^+\pi^-\eta_c$ can reach up to several tens of keV. We hope that the current calculations within the molecular interpretation will be helpful for the future experiments.
\end{abstract}

\date{\today}
\maketitle

\section{Introduction}\label{section:introduction}
In 2003, the Belle collaboration discovered the $X(3872)$, also known as $\chi_{c1}(3872)$, in the $J/\psi \pi^- \pi^+$ invariant mass spectrum of the $B\rightarrow K^+ J/\psi \pi^- \pi^+$ process~\cite{Belle:2003nnu}. The discovery of $X(3872)$ opened up a new era for the study of hadron spectroscopy. Candidates of exotic states are usually called collectively as $XYZ$ states. In the $XYZ$ states, the $X(3872)$ is the first and most extensively studied representative. Shortly, the BaBar~\cite{BaBar:2004oro,BaBar:2008qzi}, D0~\cite{D0:2004zmu}, and CDF~\cite{CDF:2003cab} collaborations confirmed its existence in $e^+e^-$ or $pp/p\bar{p}$ collisions. The spin-parity quantum numbers $J^{PC}=1^{++}$ are well established based on the angular correlation analyses in $B^+\to X(3872) K^+ \to \pi^+\pi^- J/\psi K^+\to \pi^+\pi^-\mu^+\mu^- K$ \cite{LHCb:2013kgk}.

The $X(3872)$ has two remarkable features. One is its width $\Gamma=(1.19 \pm 0.21)$ MeV~\cite{ParticleDataGroup:2024cfk}, which is exceptionally narrow compared to the typical hadronic width. Another is its mass $M_X=(3871.64 \pm 0.06) $ MeV~\cite{ParticleDataGroup:2024cfk}, which is very close to the $D^0 \bar{D}^{\ast 0}$ threshold. This near-threshold property suggests that the $X(3872)$ is a potential hadronic molecular state composed of $D\bar{D}^\ast$, which has attracted considerable attention ~\cite{Close:2003sg,Pakvasa:2003ea,Swanson:2003tb,Swanson:2004pp,Tornqvist:2004qy,Voloshin:2003nt,Wong:2003xk,AlFiky:2005jd,Braaten:2006sy,Fleming:2007rp,Ding:2009vj,Dong:2009yp,Lee:2009hy,Lee:2011rka,Liu:2009qhy,Zhang:2009vs,Gamermann:2009uq,Mehen:2011ds,Nieves:2011zz,Nieves:2012tt,Li:2012cs,Sun:2012sy,Guo:2013sya,He:2014nya,Zhao:2014gqa,Guo:2014taa,Guo:2014hqa,Braaten:2003he,Wu:2021udi,Yamaguchi:2019vea}. Guo $et$ $al$. presented a detailed review on hadronic molecules~\cite{Guo:2017jvc} and a special review focusing on the $X(3872)$ within the molecular model was recently published~\cite{Kalashnikova:2018vkv}. Generally speaking, hadronic molecules can couple to other states having the same quantum numbers, e.g., an admixture with the excited states of $c\bar{c}$~\cite{Dong:2008gb,Suzuki:2005ha}. There are other explanations of the $X(3872)$, such as the tetraquark state~\cite{Maiani:2004vq,Maiani:2005pe,Maiani:2007vr,Terasaki:2007uv,Dubnicka:2020yxy,Wang:2023sii,Wang:2019tlw} and the conventional charmonium state~\cite{Barnes:2003vb,Suzuki:2005ha}. Since no charged partner of $X(3872)$ has been observed~\cite{Belle:2011vlx,BaBar:2004cah}, the $X(3872)$ could be anticipated to be an isoscalar in the isospin symmetric limit.

In order to understand the internal structure of the $X(3872)$, a direct way is to study its various decay modes. In Ref. \cite{Dong:2009yp}, the $X(3872)$ was considered as a composite hadronic state composed of a dominant molecular $D^0D^{*0}$ component accompanied by other hadronic pairs ($D^{\pm}D^{*\pm}$, $J/\psi \omega$, and $J/\psi\rho$), with two- and three-body hadronic decays into the charmonium states $\chi_{cJ}$ plus pions being calculated. Additionally, the processes $X(3872)\to J/\psi + h$ ($h=\pi^+\pi^-,\,\pi^+\pi^-\pi^0,\,\pi^0\gamma$, and $\gamma$) were also calculated under the same framework. In Ref. \cite{Gamermann:2009fv}, the ratio of branching fractions between the $X(3872)\to J/\psi \pi^+ \pi^-\pi^0$ and $X(3872)\to J/\psi \pi^+ \pi^-$ processes was calculated to be 1.4, which is consistent with experimental measurements \cite{Belle:2005lfc}. In Ref. \cite{Guo:2014taa}, the radiative transitions of the $X(3872)$ to the $J/\psi$ and $\psi'$ were discussed using an effective field theory. It was found that the experimentally determined ratio $R = \mathcal{B}(X(3872)\to\gamma\psi')/\mathcal{B}(X(3872)\to\gamma J/\psi)$ is not in conflict with a wave function of the $X(3872)$ dominated by the $D\bar{D}^\ast$ hadronic molecular component. It is pointed out in Ref. \cite{Dong:2009uf} that, to explain the ratio $R$ mentioned above, the nontrivial interplay between a possible charmonium and the molecular components in the $X(3872)$ is necessary. However, the ratio $\mathcal{B}(X(3872)\to \gamma J/\psi)/\mathcal{B}(X(3872)\to \pi^+\pi^- J/\psi) $ relating the radiative and strong decays hints that the $c\bar{c}$ component only plays a subleading role.

Recently, the charmless decays of $X(3872)$ via the intermediate meson loops were studied based on a molecular interpretation of the $X(3872)$ as a $\bar{D}D^{*}$ bound state \cite{Wang:2022qxe}. These decays are structure-dependent, which is reflected by the proportion of neutral and charged constituents in the $X(3872)$. The dependence of certain decay widths on cutoff parameter is not significant, implying that the dominant mechanism governed by the intermediate mesonic loops effectively regulates the ultraviolet contribution. The $X(3872) \to \chi_{cJ} \pi\pi$ via intermediate mesonic box loops was calculated under the assumption that the $X(3872)$ is a shallow bound state of neutral charmed mesons~\cite{Fleming:2008yn}. They found that, at leading order, the decays of $X(3872) \to \chi_{cJ} \pi^0\pi^0$ dominate the dipionic decay modes, with a branching fraction just below that for the single pionic decay to $\pi^0 \chi_{c1}$, while the other dipionic decays are highly suppressed. In Ref. \cite{Achasov:2024anu}, the branching ratios of $X(3872)\to \pi^0 \pi^0 \chi_{c1}$ and $X(3872)\to \pi^+ \pi^- \chi_{c1}$ decay were calculated under the $D\bar{D}^*$ molecular model. The model has a significant predicted value $\mathcal{B}(X(3872)\to \pi^0 \pi^0 \chi_{c1})/\mathcal{B}(X(3872)\to \pi^+ \pi^- \chi_{c1})\approx1.1$, which is weakly dependent on the resonance parameters of $X(3872)$ and severely disrupts the isotopic symmetry.

The BaBar Collaboration obtained at the $90\%$ confidence-level upper limits on the products of the branching fractions and two-photon widths, $\Gamma_{X(3872) \to \gamma \gamma} \mathcal{B}(X(3872) \to \eta_c \pi \pi) < 11.1~\rm{eV}$~\cite{BaBar:2012mhg}, where $X(3872)$ was assumed to be a spin-2 state ($1^1D_2$ state $\eta_{c2}$) to check the $\eta_{c2}$ configuration of the $X(3872)$, but they did not observe a significant signal for this channel. To better understand the nature of $X(3872)$, in this work, under the assumption that X(3872) is a $J^{PC}=1^{++}$ molecular state, we shall study the hidden charmed decay $X(3872)\rightarrow \pi^+ \pi^- \eta_c$ via the triangle and box loops using the effective Lagrangian method within the framework of the molecular model. The influence of the neutral and charged configurations of the $X(3872)$ and its mass are also considered. The rest structure of this article is organized as follows. We provide the effective Lagrangians and the corresponding decay amplitudes in Sec.~\ref{sec:formalism}. Numerical results and discussion are given in Sec.~\ref{section:results}. The last section is devoted to a brief summary.


\section{THEORETICAL FRAMEWORK}\label{sec:formalism}
\subsection{Effective Lagrangians}\label{sec:2.1}
\begin{figure*}
	\centering
	\includegraphics[width=0.85\linewidth]{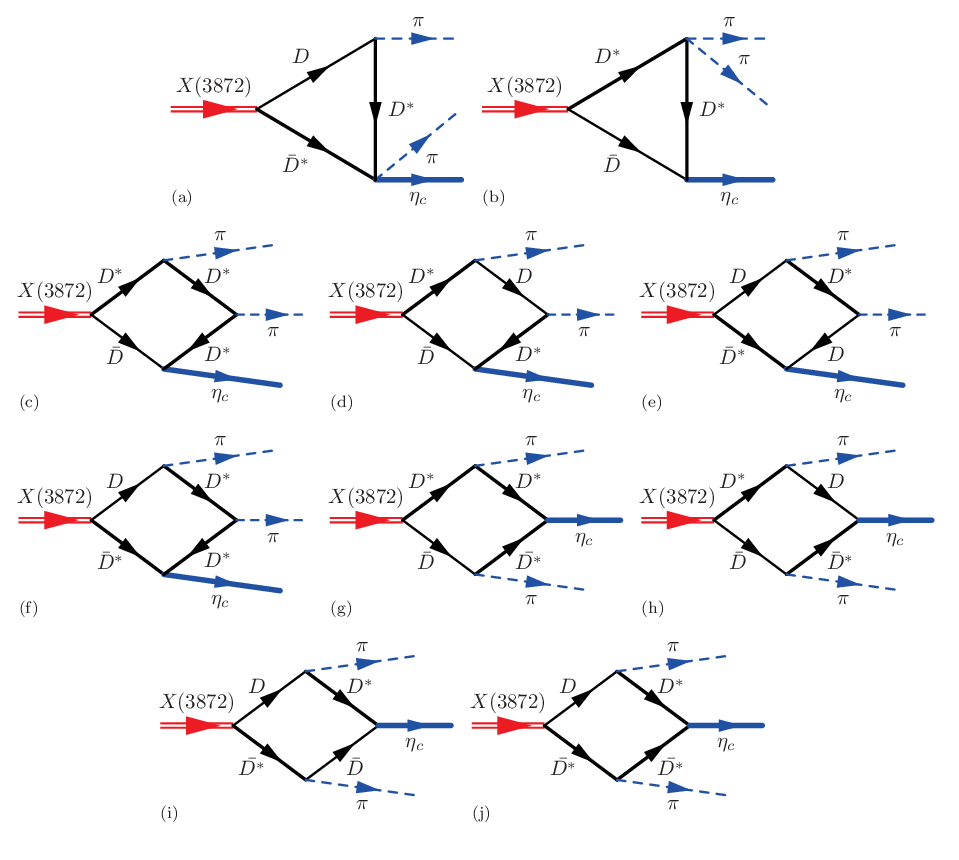}\\
    \caption{Feynman diagrams for the $X(3872) \to \pi \pi \eta_c$ decays via the triangle and box loops. The charge conjugated loops are not shown here, but included in our calculations. }
    \label{fig:feyndiags}
\end{figure*}

The process of the $X(3872)$ decaying into $\pi^+ \pi^- \eta_c$ via the intermediate charmed meson loops is shown in Fig.~\ref{fig:feyndiags}. The $X(3872)$ is assumed to be an $S$-wave molecular state with the quantum numbers of $I^G(J^{PC})=0^+(1^{++})$, consisting of a superposition of the $D^{\ast 0} \bar{D}^0$ and $D^{\ast \pm } {D}^{\mp} $ double hadron configurations~\cite{Wang:2022qxe}
\begin{eqnarray}\label{eq:x3872WF}
\left|X(3872)\right\rangle &=& \frac{\cos \theta}{\sqrt2}(D^{\ast 0} \bar{D}^0+D^0 \bar{D}^{\ast0})\nonumber\\ &+& \frac{\sin\theta}{\sqrt2}(D^{\ast + } {D}^{-}+D^+ D^{\ast -}),
\end{eqnarray}
where $\theta$ is a phase angle introduced to describe the relative proportion of the neutral and charged constituent components, and the coupling constants are sensitive to the mass difference  between the charged and neutral configurations of $X(3872)$. Here we take $\theta=0$, $\pi/12$, $\pi/4$, and $\pi/2$. The coupling between the $X(3872)$ and the charmed mesons can be parameterized using the following Lagrangian~\cite{Wang:2022qxe}
 \begin{align}\label{eq:X3872Lagrangian} 
 \mathcal{L}_{X} &= \frac{g_n}{\sqrt2}X^{\dagger}_{\mu}(D^{\ast 0 \mu} \bar{D}^0+D^0 \bar{D}^{\ast0\mu})\nonumber\\ &+ \frac{g_c}{\sqrt2}X^{\dagger}_{\mu}(D^{\ast + \mu} {D}^{-}+D^+ D^{\ast -\mu}),
\end{align}
where $g_n=|g_{\rm{eff}}^n|\cos{\theta}$ and $g_c=|g_{\rm{eff}}^c|\sin{\theta}$ represent the coupling constants of the $X(3872)$ to the neutral and charged charm meson pairs, respectively, with the effective coupling constant $g_{\rm{eff}}^{n/c}$ of the $X(3872)$ to its components.

As a state slightly below an $S$-wave two-hadron threshold, the effective coupling constant $g_{\rm{eff}}$ of the $X(3872)$ to its components is~\cite{Baru:2003qq,Weinberg:1965zz,Guo:2013zbw},
\begin{eqnarray}\label{eq:xs1} 
	g^{2}_{\rm{eff}}=16\pi c_{i}^{2}(m_{D}+m_{\bar{D}^*})^2\sqrt{\frac{2\epsilon}{\mu}}\,, 
\end{eqnarray}
where $c_{i}^2$ with $i=D^0\bar{D}^{*0}$ or $D^{\pm}\bar{D}^{*\mp}$ is the probability of finding the two-hadron component in the physical wave function of the $X(3872)$. For a pure bound state $c_{i}^2=1$~\cite{Guo:2013zbw}. $\epsilon =m_D+m_{\bar{D}^*}-M_{X}$ is the binding energy and $\mu =m_D m_{\bar{D}^*}/(m_D+m_{\bar{D}^*})$ is the reduced mass. According to the PDG~\cite{ParticleDataGroup:2024cfk} and assuming that the $X(3872)$ is a pure $D^0\bar{D}^{*0}$ or $D^\pm \bar{D}^{* \mp}$ molecule, we obtain
\begin{subequations} \label{eq:xs2}
  \begin{align}
  \left|g^n_{\rm{eff}}\right| &= 2.77~\rm{GeV} ,  \\
  \left| g^c_{\rm{eff}}\right| &= 9.95~\rm{GeV}.
  \end{align}
\end{subequations}

In the heavy quark limit, the Lagrangian for the interactions of the $S$-wave charmonium $\eta_c$ with $D$ and $D^{*}$ mesons is~\cite{Colangelo:2003sa} 
\begin{align}\label{eq:etacDD} 
 \mathcal{L}_S &=-g_{\eta_c D^\ast D^\ast} \epsilon_{\mu \nu \alpha \beta}\partial^\mu \eta_{c}^\dagger D^{\ast \nu} \pararrowk{\alpha} \bar{D}^{\ast \beta}\nonumber\\ &+
 \ii g_{\eta_{c}D^\ast D} \eta_{c}^{\dagger}(D\pararrowk{\mu}\bar{D}^\ast_\mu+D^\ast_\mu \pararrowk{\mu} \bar{D}) +\mathrm{H.c.}\,.
\end{align}
The $D^{(\ast)}=(D^{(\ast) 0 }, D^{(\ast) +}, D^{(\ast) +}_s)$ represents the pseudoscalar (vector) triplet of charmed mesons. The corresponding anticharmed meson triplets are denoted by $\bar{D}^{(\ast)}$. The coupling constants $g_{\eta_{c} D^{(\ast)} D^{(\ast)}}$ can be written as
 \begin{subequations} \label{eq:xs5}
	\begin{align}
		g_{\eta_c D^\ast D^\ast} &= 2g_1 \frac {m_{D^*}}{\sqrt{m_{\eta_{c}}}} \,,\\ 
		g_{\eta_{c}D^\ast D} &= 2g_1\sqrt{m_{\eta_{c}} m_{D^{\ast}} m_D} \,,
	\end{align}
\end{subequations}
where the gauge coupling $g_1=\sqrt{m_\psi}/(2m_D f_\psi)$ with the $J/\psi$ decay constant  $f_\psi=426~\rm{MeV}$~\cite{Wu:2021udi,Colangelo:2003sa}.

In the heavy quark limit and under the chiral symmetry, the interactions of the light pseudoscalar mesons with $D$ and $D^{*}$ mesons are described by the following Lagrangian~\cite{Wu:2022hck,Zheng:2024eia,Grinstein:1992qt}
\begin{align}\label{eq:PDD}
	\mathcal{L} &= -\ii g_{D^*DP}\big(D^{i }\partial^{\mu} P_{ij} D_\mu^{*j\dagger} - D_\mu^{*i}\partial^\mu P_{ij} D^{j\dagger}) \nonumber\\
	&+ \frac{1}{2} g_{D^*D^*P}\epsilon_{\mu\nu\alpha\beta} D_i^{*\mu}\partial^\nu P^{ij}\pararrowk{\alpha} D_j^{*\beta \dagger}+\mathrm{H.c.}\,,
\end{align}
where $P$ represents the octet pseudoscalar  mesons in matrix form~\cite{Wang:2022qxe} 
\begin{eqnarray}\label{eq:matrix}
P&=\begin{pmatrix}
	\frac{\pi^0}{\sqrt{2}} + \frac{\eta}{\sqrt{6}} & \pi^+ & K^+\\
	\pi^- & -\frac{\pi^0}{\sqrt{2}} + \frac{\eta }{\sqrt{6}} & K^0\\
	K^- & \bar{K}_0 & -\sqrt {\frac{2 }{3}}\eta \label{eq:P}
\end{pmatrix}\,.
\end{eqnarray}
The coupling constants ${g_{D^{(\ast)}D^{(\ast)}P}}$ are
\begin{align}\label{eq:xs6}
g_{D^{\ast}D^{\ast}P}&=\frac{g_{D^{\ast}DP}}{\sqrt{m_D m_{D^\ast}}}=\frac{2g}{f_\pi}\,,
\end{align}
where the pion decay constant $f_\pi=132~\rm{MeV}$~\cite{Casalbuoni:1996pg} and $g = 0.59$ which is determined from the partial decay widths of $D^{*+}\to D^0\pi^+/D^+\pi^0$ ~\cite{CLEO:2001foe}.

The Lagrangian describing the four-body interaction between $\eta_c$, pions and $D^*$ mesons, which is constructed via the $SU(4)$ matrix of pseudoscalar and vector mesons, is~\cite{Lin:1999ad,Haglin:1999xs}
\begin{align}\label{eq:etacpiDD} 
 \mathcal{L}_{4B} &=g_{\eta_c \pi D^\ast D^\ast}\eta_c D^{\ast 0}_\mu D^{\ast  - \mu}\pi^+ \nonumber\\&+g_{\eta_c \pi D^\ast D^\ast}\eta_c \bar{D}^{\ast 0 \mu} D^{\ast+ }_\mu \pi^- \nonumber\\
&+ g_{\pi \pi D^\ast D^\ast}\pi^-D^{\ast -}_\mu \pi^+ D^{\ast + \mu} \nonumber\\
&+ g_{\pi \pi D^\ast D^\ast}\pi^-D^{\ast 0}_\mu\pi^+\bar{D}^{\ast 0 \mu}+\mathrm{H.c.}\,,
\end{align}
with the coupling constants $g_{\eta_{c}\pi D^{\ast} D^{\ast}}=32.26$ and $g_{\pi \pi D^\ast D^\ast}=13.97$.

\subsection{Amplitudes of $X (3872) \rightarrow \pi^+ \pi^- \eta_ {c}$}\label{sec:2.2}

Using the Lagrangians above, we can write down the amplitudes of $X (3872) \rightarrow \pi^+ \pi^- \eta_ {c}$ shown in Fig.~\ref{fig:feyndiags}. The amplitudes of the process $X(3872)(p)\rightarrow[D^{(*)}(q_1)\bar{D}^{(*)}(q_2)]D^{(*)}(q_3) \rightarrow  \pi^- (p_1) \pi^+ (p_2) \eta_c (p_3)$ via the triangle loops are
\begin{eqnarray}\label{eq:triangle1} 
\mathcal{M}_a &=&\int \frac{d^4 q_3}{(2\pi)^4}\left[\frac{g_{\rm{eff}}}{\sqrt{2}} g_{\mu \gamma}\varepsilon^\mu(X)\right] \left[ g_{D^\ast D P} p_{1 \nu}\right] \nonumber\\ &\times&\left[g_{D^\ast D^\ast \eta_{c} \pi} g_{\beta \alpha}\right]S(q_1,m_{D})S^{\gamma \beta}(q_2,m_{D^\ast})\nonumber\\ &\times&S^{\alpha \nu}(q_3,m_{D^\ast})F^{2}(q_3^2,m^2_{D^*})\,,
\end{eqnarray}
\begin{eqnarray}\label{eq:triangle2} 
	\mathcal{M}_b &=&\int \frac{d^4 q_3}{(2\pi)^4}\left[\frac{g_{\rm{eff}}}{\sqrt{2}} g_{\mu \theta}\varepsilon^\mu(X)\right]\left[g_{D^\ast D^\ast \pi \pi} g_{\rho \xi}\right]\nonumber\\ &\times&\left[ g_{D^\ast D\eta_{c}}(q_2-q_3)^\delta g_{\delta \zeta}\right]S^{\theta  \rho}(q_1,m_{D^\ast})\nonumber\\ &\times&S(q_2,m_{D})S^{\xi \zeta}(q_3,m_{D^\ast})F^{2}(q_3^2,m^2_{D^*})\,.
\end{eqnarray}
The amplitudes of the $X(3872)(p)\rightarrow [D^{(*)}(q_1) \bar{D}^{(*)}(q_2)]$ $ D^{(*)}(q_3) D^{(*)}(q_4) \rightarrow \pi^- (p_1) \pi^+ (p_2) \eta_c (p_3)$ via the box loops are
\begin{eqnarray}\label{eq:zhenfu1} 
\mathcal{M}_c &=&\int \frac{d^4 q_4}{(2\pi)^4}\left[\frac{g_{\rm{eff}}}{\sqrt{2}} g_{\mu \gamma}\varepsilon^\mu(X)\right]\nonumber\\ &\times&\left[-\frac{1}{2}g_{D^\ast D^\ast P}\epsilon_{\delta  \nu \alpha  \beta } p_{1}^\nu  (q_1+q_4)^\alpha\right]\nonumber\\ &\times&\left[-\frac{1}{2}g_{D^\ast D^\ast P}\epsilon_{\omega \sigma \theta  \iota } p_{2}^\sigma (q_4+q_3)^\theta\right]\left[g_{D^\ast D \eta_{c}}(q_2-q_3)_\kappa\right]\nonumber\\&\times&S^{\gamma \delta}(q_1,m_{D^\ast})S(q_2,m_{D})S^{\iota \kappa}(q_3,m_{D^\ast})\nonumber\\ &\times&S^{\beta \omega}(q_4,m_{D^\ast})F(q_3^2,m^2_{D^*})F(q_4^2,m^2_{D^*})\,,
\end{eqnarray}
\begin{eqnarray}\label{eq:zhenfu2} 
\mathcal{M}_d &=&\int \frac{d^4 q_4}{(2\pi)^4}\left[\frac{g_{\rm{eff}}}{\sqrt{2}} g_{\mu \gamma}\varepsilon^\mu(X)\right]\left[-g_{D^\ast D P} p_{1 \delta}\right]\left[ g_{D^\ast D P}p_{2 \iota}\right]\nonumber\\&\times&\left[ g_{D^\ast D \eta_{c}}(q_2-q_3)^\kappa\right] S^{\gamma \delta}(q_1,m_{D^\ast})S(q_2,m_{D})\nonumber\\&\times&S^{\iota \kappa}(q_3,m_{D^\ast})S(q_4,m_{D})\nonumber\\&\times&F(q_3^2,m^2_{D^*})F(q_4^2,m^2_D)\,,
\end{eqnarray}
\begin{eqnarray}\label{eq:zhenfu3} 
\mathcal{M}_e &=&\int \frac{d^4 q_4}{(2\pi)^4}\left[\frac{g_{\rm{eff}}}{\sqrt{2}} g_{\mu \xi}\varepsilon^\mu(X)\right]\left[ g_{D^\ast D P} p_{1 \beta}\right]\left[ - g_{D^\ast DP}p_{2 \omega}\right]\nonumber\\&\times&\left[ g_{D^\ast D \eta_{c}}(q_2-q_3)_\rho \right] S(q_1,m_{D}) S^{\xi \rho}(q_2,m_{D^\ast})\nonumber\\&\times&S(q_3,m_{D})S^{\beta \omega}(q_4,m_{D^\ast})\nonumber\\&\times&F(q_3^2,m^2_D)F(q_4^2,m^2_{D^*})\,,
\end{eqnarray}
\begin{eqnarray}\label{eq:zhenfu4}
\mathcal{M}_f &=&\int \frac{d^4 q_4}{(2\pi)^4}\left[\frac{g_{\rm{eff}}}{\sqrt{2}} g_{\mu \xi}\varepsilon^\mu(X)\right]\left[ g_{D^\ast D P} p_{1 \beta}\right]\nonumber\\ &\times&\left[-\frac{1}{2}g_{D^\ast D^\ast P}\epsilon_{\omega \delta \theta  \iota }p_{2}^\delta (q_4+q_3)^\theta\right]\nonumber\\ &\times&\left[-g_{D^\ast D^\ast \eta_{c}}\epsilon_{\sigma \kappa \tau \rho} p_3^\sigma (q_2-q_3)^\tau\right]\nonumber\\ &\times& S(q_1,m_{D}) S^{\xi \rho}(q_2,m_{D^\ast})S^{\iota \kappa}(q_3,m_{D^\ast})\nonumber\\&\times&S^{\beta \omega}(q_4,m_{D^\ast})F(q_3^2,m^2_{D^*})F(q_4^2,m^2_{D^*})\,,
\end{eqnarray}
\begin{eqnarray}\label{eq:zhenfu5} 
	\mathcal{M}_g &=&\int \frac{d^4 q_4}{(2\pi)^4}\left[\frac{g_{\rm{eff}}}{\sqrt{2}} g_{\mu \gamma}\varepsilon^\mu(X)\right]\nonumber\\ &\times&\left[-\frac{1}{2}g_{D^\ast D^\ast P}\epsilon_{\delta  \nu \alpha  \beta } p_{1}^\nu (q_1+q_4)^\alpha \right]\nonumber\\ &\times&\left[- g_{D^\ast D  P} p_{2 \iota}\right]\left[-g_{D^\ast D^\ast \eta_{c}}\epsilon_{\sigma \kappa \tau \rho} p_3^\sigma  (q_3-q_4)^\tau\right]\nonumber\\&\times&S^{\gamma \delta}(q_1,m_{D^\ast})S(q_2,m_{D})S^{\iota \rho}(q_3,m_{D^\ast})\nonumber\\ &\times&S^{\beta \kappa}(q_4,m_{D^\ast})F(q_3^2,m^2_{D^*})F(q_4^2,m^2_{D^*})\,,
\end{eqnarray}
\begin{eqnarray}\label{eq:zhenfu6} 
	\mathcal{M}_h &=&\int \frac{d^4 q_4}{(2\pi)^4}\left[\frac{g_{\rm{eff}}}{\sqrt{2}} g_{\mu \gamma}\varepsilon^\mu(X)\right]\left[- g_{D^\ast D P} p_{1 \delta}\right]\nonumber\\&\times&\left[ - g_{D^\ast D P}p_{2 \iota}\right]\left[ g_{D^\ast D \eta_{c}}(q_3-q_4)_\rho\right] S^{\gamma \delta}(q_1,m_{D^\ast})\nonumber\\&\times&S(q_2,m_{D})S^{\iota \rho}(q_3,m_{D^\ast})\nonumber\\&\times&S(q_4,m_{D})F(q_3^2,m^2_{D^*})F(q_4^2,m^2_D)\,,
\end{eqnarray}
\begin{eqnarray}\label{eq:zhenfu7} 
\mathcal{M}_i &=&\int \frac{d^4 q_4}{(2\pi)^4}\left[\frac{g_{\rm{eff}}}{\sqrt{2}} g_{\mu \xi}\varepsilon^\mu(X)\right]\left[ g_{D^\ast D P} p_{1 \beta}\right]\left[ g_{D^\ast D P}p_{2 \omega}\right]\nonumber\\&\times&\left[ g_{D^\ast D \eta_{c}}(q_3-q_4)_\kappa\right] S(q_1,m_{D}) S^{\xi \omega}(q_2,m_{D^\ast})\nonumber\\&\times&S(q_3,m_{D})S^{\beta \kappa}(q_4,m_{D^\ast})\nonumber\\&\times&F(q_3^2,m^2_D)F(q_4^2,m^2_{D^*})\,,
\end{eqnarray}
\begin{eqnarray}\label{eq:zhenfu8} 
	\mathcal{M}_j &=&\int \frac{d^4 q_4}{(2\pi)^4}\left[\frac{g_{\rm{eff}}}{\sqrt{2}} g_{\mu \xi}\varepsilon^\mu(X)\right]\left[ g_{D^\ast D P}p_{1 \beta}\right]\nonumber\\ &\times&\left[ \frac{1}{2}g_{D^\ast D^\ast P}\epsilon_{\iota  \delta \theta  \omega }p_{2}^\delta  (q_2+q_3)^\theta\right]\nonumber\\ &\times& \left[-g_{D^\ast D^\ast \eta_{c}}\epsilon_{\sigma \kappa \tau \rho} p_3^\sigma (q_3-q_4)^\tau\right]\nonumber\\ &\times& S(q_1,m_{D}) S^{\xi \omega}(q_2,m_{D^\ast})\nonumber\\ &\times&S^{\rho \iota}(q_3,m_{D^\ast})S^{\beta \kappa}(q_4,m_{D^\ast})\nonumber\\&\times&F(q_3^2,m^2_{D^*})F(q_4^2,m^2_{D^*})\,.
\end{eqnarray}
Here, $S(q,m)$ and $S^{\mu \nu}(q,m)$ represent the propagators of the charm meson $D$ and $D^\ast$, respectively,
\begin{subequations} \label{eq:xs7}
	\begin{align}
	S(q,m_D)&=\frac{1}{q^2-m_D^2+\ii \epsilon}\,,\\
	S^{\mu \nu}(q,m_{D^\ast})&=\frac{-g^{\mu \nu}+q^\mu q^\nu/m^2_{D^\ast}}{{q^2-m_{D^\ast}^2+\ii \epsilon}}\,.
	\end{align}
\end{subequations}

The differential decay width for the $X (3872) \rightarrow \pi^+ \pi^- \eta_ {c}$ is expressed as follows
\begin{eqnarray}\label{eq:width}
	d\Gamma=\frac{1}{3}\frac{1}{(2\pi)^3}\frac{1}{32M_X^3}\left\vert {\mathcal{M}}\right\vert^2 dm_{12}^2 dm_{23}^2\,.
\end{eqnarray}
For the triangle diagrams, $\mathcal{M}=\mathcal{M}_a+\mathcal{M}_b$, and $\mathcal{M}=\sum_{i=c}^{j}\mathcal{M}_i$ for the box diagrams. In this work, the total amplitude for each diagram needs to account for both the neutral and charged components of $X(3872)$, as well as the charge conjugated for each diagram. The $m_{12}$ and $m_{23}$ are the invariant masses of $\pi^- \pi^+ $ and $\pi^+ \eta_c$, respectively.
The calculations were conducted with the assistance of LoopTools packages~\cite{Hahn:1998yk,Denner:1991kt}.
 
\section{NUMERICAL RESULTS}\label{section:results}

To describe the off-shell effects of the exchanged mesons and to account for the internal structure of the involved mesons, we introduce a form factor $F(q^2, m^2)$. In this work, the dipole form factor is adopted in triangle diagrams and two monopole form factors are adopted in box diagrams.\cite{Cheng:2004ru,Zheng:2024eia,Wang:2022qxe,Wu:2016ypc,Guo:2016iej},
\begin{eqnarray}\label{eq:xzyz}
F(q^2,m^2)=\frac{m^2-\Lambda^2}{q^2-\Lambda^2}\,,
\end{eqnarray}
where $q$ and $m$ are the momentum and mass of the exchanged meson, respectively; $\Lambda=m+\alpha \Lambda_{QCD}$ with $\Lambda_{QCD}=0.22~\rm{GeV}$. We vary the model parameter $\alpha$ within the range of $0.6$ to $1.8$ to show its effect on the decay process under consideration.

The phase angle $\theta$ quantifies the relative contributions of the neutral ($D^0 {\bar D}^{*0}$) and charged ($D^{\pm} {\bar D}^{*\mp}$) components in the $X(3872)$ molecular wave function. The sensitivity of the coupling constants $g_{n/c}$ in Eq.~(2) to the mass difference between these configurations arises from that the binding energy  $\epsilon =m_D+m_{\bar{D}^*}-M_{X}$ depends on the $X(3872)$ mass $M_{X}$. For example, if $M_{X}$ approaches the $D^0 {\bar D}^{*0}$  threshold ( $\sim 3871.7$ MeV), the neutral component dominates ($\theta\to 0$), while proximity to the charged threshold ($\sim 3879.9$ MeV) favors $\theta\to \pi/2$. In the hadronic loop approach, the charged and neutral charmed meson loops cancel each other in isospin-violating processes~\cite{Mehen:2015efa,Zhang:2024fxy,Wu:2021udi}. Therefore, one can determine the range of the phase angle $\theta$ by using experimentally observed isospin-violating processes $X(3872)\to \chi_{cJ}\pi^0$, $J/\psi\rho^0$~\cite{ParticleDataGroup:2024cfk}. In the following, we shall present the decay widths of $X(3872)\rightarrow \pi^+ \pi^- \eta_{c}$ under four different phase angles: $\theta=0$, $\pi/12$, $\pi/4$, and $\pi/2$. When $\theta=0$, the $X(3872)$ is a purely neutral bound state, consisting solely of the neutral component. For $\theta=\pi/12$, the decay process involves both neutral and charged components, but the neutral component dominates. For $\theta=\pi/4$, the charged and neutral components of $X(3872)$ have equal proportions. At $\theta=\pi/2$, the $X(3872)$ is a purely charged bound state, consisting solely of a charged component. 

\begin{figure}
	\centering
	\includegraphics[width=0.95\linewidth]{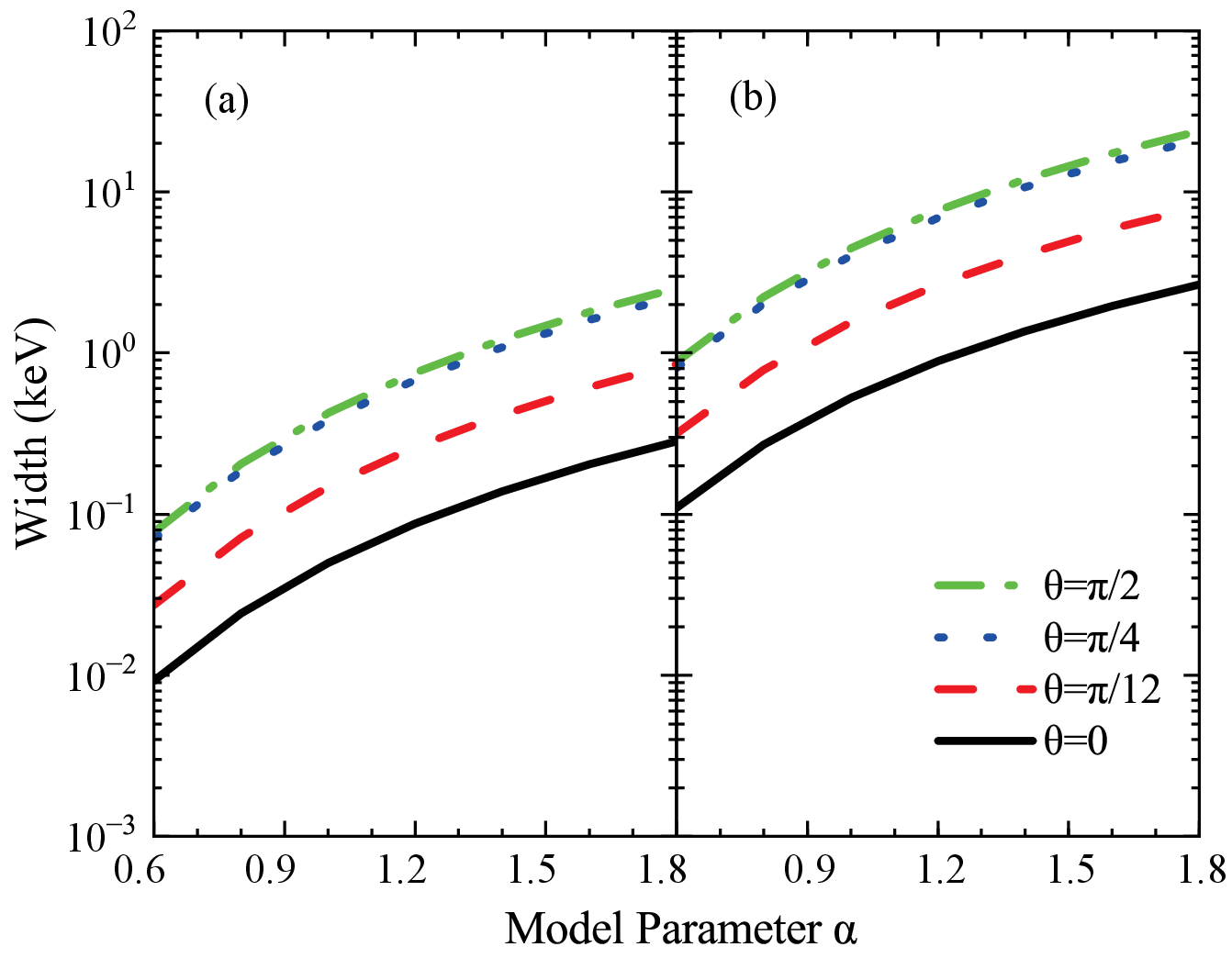}
	\caption{Partial widths (in units of keV) of the decay $X(3872)\rightarrow \pi^{+}  \pi^{-} \eta_{c}$ through triangle loops (a) and box loops (b) as a function of the model parameter $\alpha$.}
	\label{fig:34P}
\end{figure}

In Fig.~\ref{fig:34P} the partial widths of the $X(3872)\rightarrow \pi^{+}  \pi^{-} \eta_{c}$ are plotted as a function of the model parameter $\alpha$. The results in the left-hand column are for the triangle loops and those due to the box loops are shown in the right-hand column. Overall, the partial decay widths contributed from the box loops are one order of magnitude larger than those from triangle ones. It can be seen that for different phase angles, the decay widths via the triangle and box loops both increase monotonously as the parameter $\alpha$ varies from 0.6 to 1.8.

\begin{table*}[htbp]
\caption{Decay widths (in units of keV) of the $X(3872)\rightarrow \pi^{+}  \pi^{-} \eta_{c}$ with $\alpha=0.6–1.8$. For better comparison, decay widths (in units of keV) of the $X(3872)\rightarrow \pi^+ \pi^- J/\psi$ ~\cite{Meng:2007cx,Ferretti:2018tco}, $ \pi^+ \pi^- \chi_{c1}$~\cite{Meng:2007cx,BESIII:2023eeb} and $ \pi^+ \pi^- \chi_{c0}$~\cite{Meng:2007cx,BESIII:2022kow} are also listed together with the experimental measurements ~\cite{ParticleDataGroup:2024cfk}. }
	\label{tab:results}
	\begin{ruledtabular}
		\begin{tabular}{lccccc}
			Final states	&$\theta = 0$& $\theta = \pi/12$&$\theta = \pi/4$&$\theta = \pi/2$ &Exp. data\footnote{The experimental data are obtained through multiplying the center value of the $X(3872)$ total width, $\Gamma_{X(3872)}=1.19\pm0.21~\rm{MeV}$~\cite{ParticleDataGroup:2024cfk}, by the branching fraction of each decay channel.}\\
			\colrule
			$\pi^{+}  \pi^{-} \eta_{c}$ (Triangle) & $0.01\sim 0.28$ & 
            $0.03\sim0.86$ &
            $0.07\sim2.25$ &
		  $0.08\sim2.53$ & $<154.7$~\cite{ParticleDataGroup:2024cfk} \\
		    $\pi^{+}  \pi^{-} \eta_{c}$ (Box)  & $0.11\sim2.66$ &
		  $0.32\sim8.04$ & 
		  $0.81\sim21.16$ &
		  $0.88\sim 23.79$ &$<154.7$~\cite{ParticleDataGroup:2024cfk}  \\
	$ \pi^+ \pi^- J/\psi$
         &$10.0\footnote{$\Gamma_{X(3872) \to  \pi^+\pi^- J/\psi}=\Gamma_{X(3872) \to \rho^0 J/\psi} \times \mathcal{B}[\rho^0 \to \pi^+ \pi^-]$, with $\Gamma_{X(3872) \to  \rho^0 J/\psi}=10~\rm{keV}$ which is independent of $\alpha$~\cite{Ferretti:2018tco} and $\mathcal{B}[\rho^0 \to \pi^+ \pi^-] = 100\%$~\cite{ParticleDataGroup:2024cfk}.}$~\cite{Ferretti:2018tco}
          & $\cdots$
           & $35.0\sim70.0$\footnote{$\Gamma_{X(3872)\to \rho^0 J/\psi}=35 \sim 70~\rm{keV}$ with $\alpha=4.0$, where the $X(3872)$, with a mass ranging from 3870.5 to 3872.5 MeV, decays through triangle diagrams~\cite{Meng:2007cx}.} ~\cite{Meng:2007cx}
            & $\cdots$
            &$41.7 \pm10.7 $~\cite{ParticleDataGroup:2024cfk}\\
        $ \pi^+ \pi^- \chi_{c1}$
        & $\cdots$ 
        & $\cdots$ 
        &  $<12.6$\footnote{Obtained by $\Gamma_{X(3872) \to \pi^+\pi^- J/\psi } \times {\mathcal{B}[X(3872) \to \pi^+ \pi^- \chi_{c1}]}/{\mathcal{B}[X(3872) \to \pi^+ \pi^- J/\psi]}$ with $\Gamma_{X(3872) \to \pi^+\pi^- J/\psi} < 70.0 ~\rm{keV}$~\cite{Meng:2007cx} and ${\mathcal{B}[X(3872) \to \pi^+ \pi^- \chi_{c1}]}/{\mathcal{B}[X(3872) \to \pi^+ \pi^- J/\psi]}<0.18$~\cite{BESIII:2023eeb}.}~\cite{Meng:2007cx}
        & $\cdots$  & $<8.3$~\cite{ParticleDataGroup:2024cfk} \\
        $ \pi^+ \pi^- \chi_{c0}$&
         $\cdots$
        & $\cdots$
        & $<39.2$\footnote{Obtained by $\Gamma_{X(3872) \to \pi^+\pi^- J/\psi} \times {\mathcal{B}[X(3872) \to \pi^+ \pi^- \chi_{c0}]}/{\mathcal{B}[X(3872) \to \pi^+ \pi^- J/\psi]}$ with $\Gamma_{X(3872) \to \pi^+\pi^- J/\psi } < 70.0 ~\rm{keV}$~\cite{Meng:2007cx} and ${\mathcal{B}[X(3872) \to \pi^+ \pi^- \chi_{c0} ]}/{\mathcal{B}[X(3872) \to \pi^+ \pi^- J/\psi ]}<0.56$~\cite{BESIII:2022kow}.}~\cite{Meng:2007cx}
        & $\cdots$ & $<23.8$~\cite{ParticleDataGroup:2024cfk} \\
		\end{tabular}
	\end{ruledtabular}
\end{table*}

In Table \ref{tab:results} we summarize the decay widths for the four different phase angles. Moreover, the other dipionic decays of the $X(3872)$ to the charmonia $J/\psi$ and $\chi_{c0(c1)}$ are also listed together with the experimental results. With the model parameters we used, the decay widths for the $X(3872)\to\pi^+\pi^-\eta_c$ can reach several tens of keV, smaller than the experimental upper limit of 154.7 keV \cite{ParticleDataGroup:2024cfk}. However, we cannot limit the range of the model parameters (phase angles $\theta$ and cutoff $\alpha$) according to the comparison between the theoretical predictions and experimental measurements.

The relative size of the decay widths for the different phase angles can be estimated in terms of the effective couplings in Eq. \eqref{eq:xs2}. 
The decay width for $\theta = \pi/2$ is roughly one order of magnitude greater than that for $\theta = 0$, as expected in view of $g^n_{\rm{eff}}/g^c_{\rm{eff}} = 3.59$. In the case of $\theta = \pi/4$, the decay width is slightly smaller than that for $\theta=\pi/2$. This is due to $|g^n_{\rm{eff}}|\cos \theta + |g^c_{\rm{eff}}|\sin \theta = 8.99 \approx g^c_{\rm{eff}}$. Given $|g^n_{\rm{eff}}|\cos \theta + |g^c_{\rm{eff}}|\sin \theta \approx 5.25$, the decay width for $\theta = \pi/12$ is about four times greater than that for $\theta=0$.

\begin{figure}
	\centering
	\includegraphics[width=0.95\linewidth]{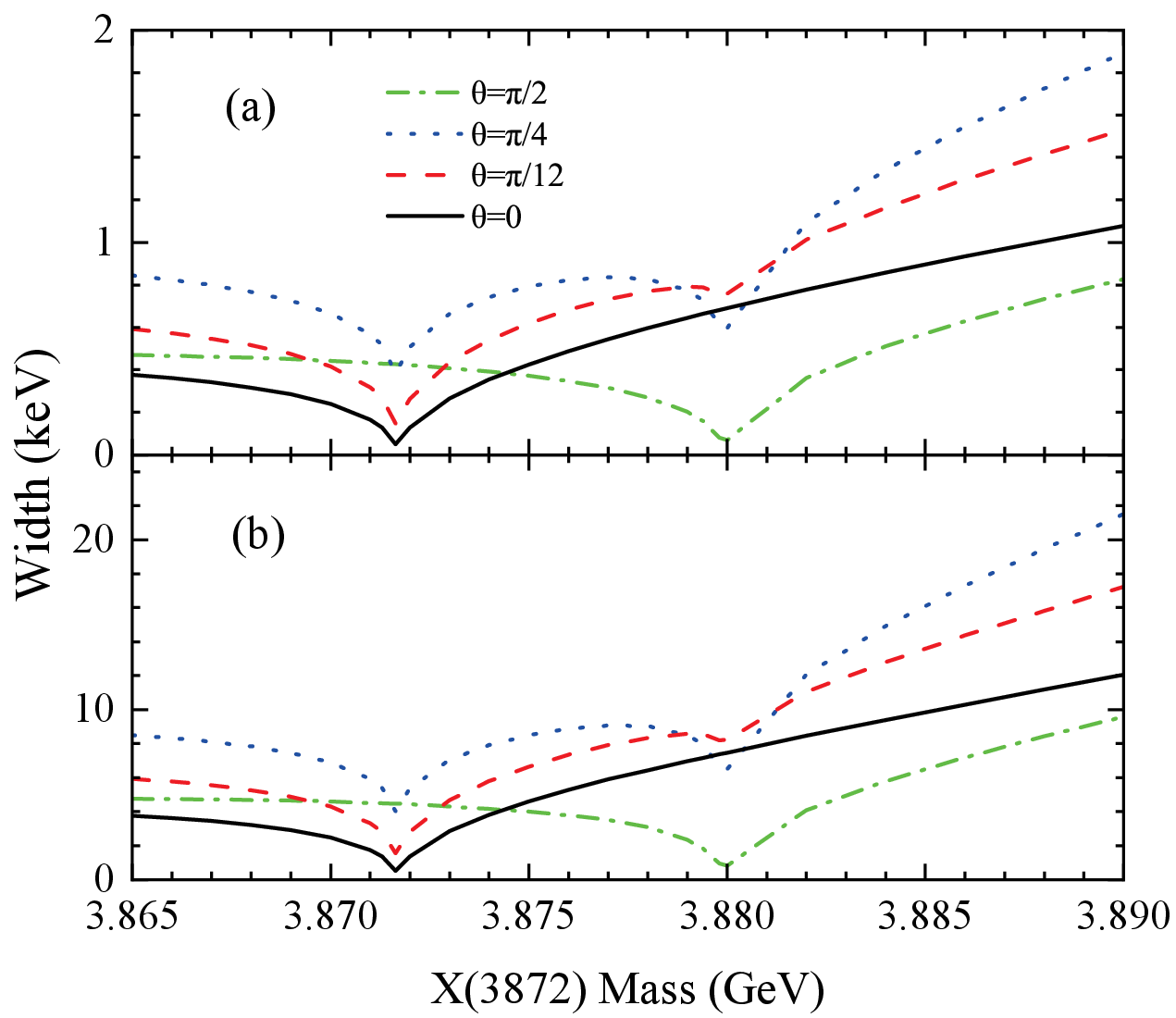}
	\caption{Partial widths for the decay $X(3872)\rightarrow \pi^+ 
	\pi^-\eta_{c}$ as a function of the mass of the $X(3872)$. The upper panel is for the triangle loops and the lower panel represents the results due to the box ones. The model parameter $\alpha=1.0$.}
	\label{fig:3P2}   
\end{figure}

It is shown in Fig.~\ref{fig:3P2} that the decay widths show dips near $M_X=3872$ MeV and $3880$ MeV for $\theta = \pi/12 $ and $\pi/4$. This is because, when the $X(3872)$ mass is close to the $D^0\bar{D}^{*0}$ and $D^{\pm}\bar{D}^{*\mp}$ threshold, the corresponding coupling strength approaches zero. At $\theta = 0$ and $\pi/2$, the $X(3872)$ is composed either of neutral charmed mesons or of charged ones, so there is one dip at $M_X=3872$ MeV or $3880$ MeV.
\begin{figure}
	\centering
	\includegraphics[width=0.95\linewidth]{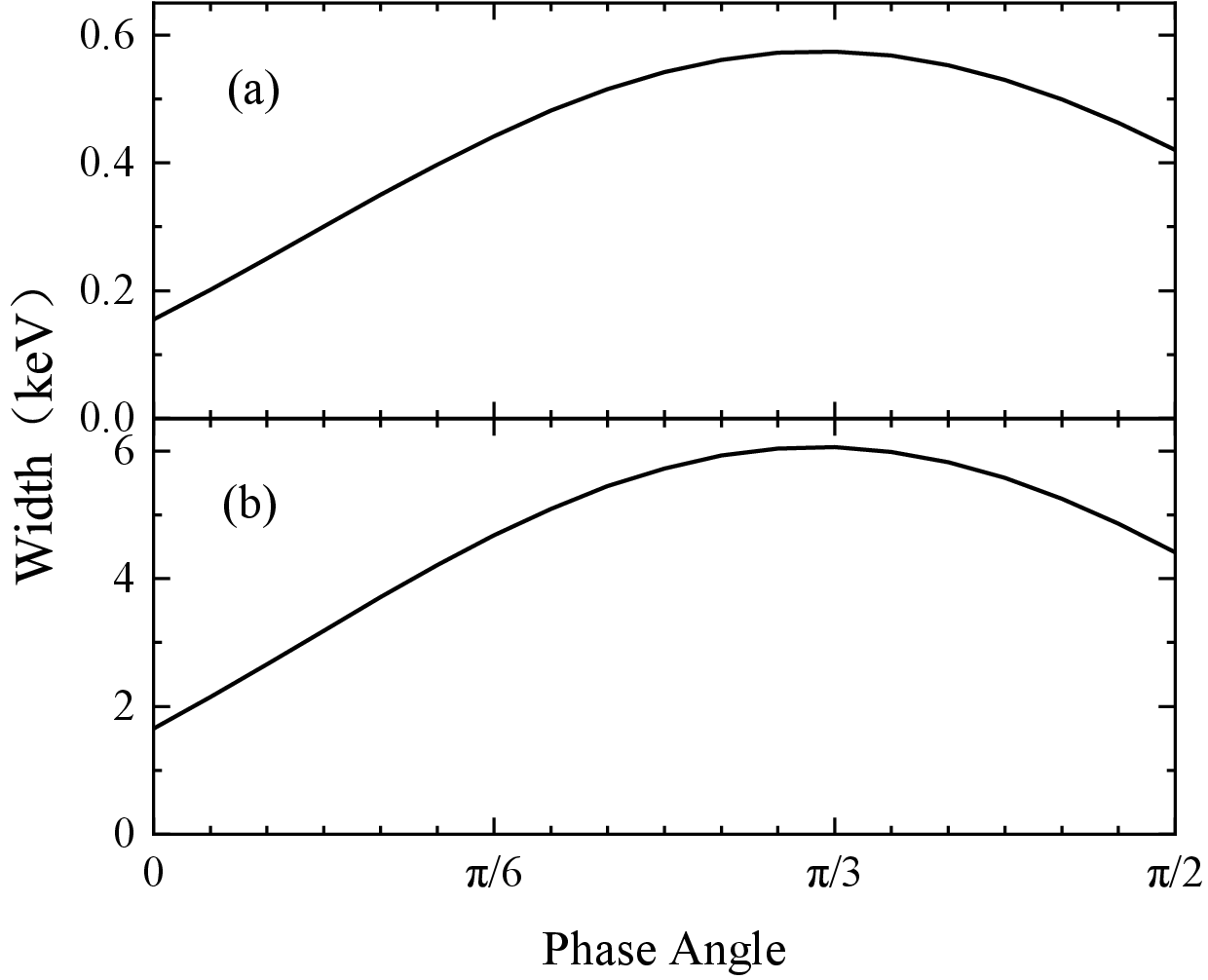}
	\caption{Partial widths for the decay $X(3872)\rightarrow \pi^+ 
	\pi^-\eta_{c}$ as a function of the phase angle. The upper panel is for the triangle loops and the lower panel represents the results due to the box ones. The model parameter $\alpha=1.0$.}
	\label{fig:3P3}   
\end{figure}

To clearly show the influence of the phase angle, we plot in Fig.~\ref{fig:3P3} the dependence of the decay width on the phase angle for $\alpha=1.0$. The upper panel presents the contributions from the triangle loops, while the lower panel exhibits those due to the box ones. For $X(3872)\rightarrow \pi^+ \pi^-\eta_{c}$, the decay width first increases and then decreases as the phase angle increases. It is seen that the contribution from the box loop is about one order of magnitude larger than those from the triangle loops. From Fig.~\ref{fig:3P3}, the decay width reaches the maximum values around $\pi/3$. According to our calculations, the obtained decay widths due to the triangle loops and box loops are about less than $1$ keV, and a few keVs, respectively. The future experimental measurements may be used to constrain the range of the phase angle.

\section{SUMMARY}\label{section:summary}

In this work, we have calculated the decay widths of the dipionic decay $X(3872)\rightarrow \pi^+ \pi^- \eta_{c}$ using an effective Lagrangian approach. In the calculations, we assumed that the $X(3872)$ is a $D{\bar D}^\ast$ molecular state and decays via the triangle and box loops. To show the influence of the $X(3872)$ molecular structure, we chose four phase angles $\theta = 0,\,\pi/12,\,\pi/4$, and $\pi/2$, which describes the proportion of neutral and charged constituents in the $X(3872)$. When $\theta=0$, the $X(3872)$ is a purely neutral bound state, consisting solely of the neutral component. For $\theta=\pi/12$, the decay process involves both neutral and charged components, but the neutral component dominates. For $\theta=\pi/4$, the charged and neutral components of $X(3872)$ have equal proportions. At $\theta=\pi/2$, the $X(3872)$ is a purely charged bound state, consisting solely of a charged component.

The calculated results indicate that the partial decay widths contributed from the box loops are one order of magnitude greater than those from the triangle loops. With the present model parameters, the decay widths can reach several tens of keV, smaller than the experimental upper limit $154.7~\mathrm{keV}$. We make a simple comparison of our calculations with other dipionic transitions of the $X(3872)$ to the charmonia $J/\psi$ or $\chi_{cJ}$.
The future precise measurements on this process could help us restrict the model parameters. 

\begin{acknowledgments}\label{sec:acknowledgements}
This work is partly supported by the National Natural Science Foundation of China under Grant Nos. 12475081, 12105153, and 12075133, and by the Natural Science Foundation of Shandong Province under Grant Nos. ZR2021MA082 and ZR2022ZD26. It is also supported by Taishan Scholar Project of Shandong Province (Grant No.tsqn202103062).	
\end{acknowledgments}

\bibliography{X3872decay}
\end{document}